\documentclass[12pt]{iopart}
\usepackage{amssymb,graphicx}

\begin{document}
	
\title[Multipole Traps as Tools in Environmental Studies]{Multipole Traps as Tools in Environmental Studies}
	
\author{Bogdan M. Mihalcea$^1$, Cristina Stan$^2$, Liviu C. Giurgiu$^3$, Andreea Groza$^1$, Agavni Surmeian$^1$, Mihai Ganciu$^1$, Vladimir E. Filinov$^4$, Dmitry Lapitsky$^4$, Lidiya Deputatova$^4$, Leonid Vasilyak$^4$, Vladimir Pecherkin$^4$, Vladimir Vladimirov$^4$, Roman Syrovatka$^4$}
\address{$^1$ National Institute for Laser, Plasma and Radiation Physics (INFLPR), Atomi\c stilor Str. Nr. 409, 077125 M\u agurele, Romania}
\address{$^2$ Department of Physics, {\em Politehnica} University, 313 Splaiul Independen\c tei, RO-060042, Bucharest, Romania} 
\address{$^3$ University of Bucharest, Faculty of Physics, Atomistilor Str. Nr. 405, 077125 M\u agurele, Romania}
\address{$^4$Joint Institute for High Temperatures, Russian Academy of Sciences, Izhorskaya Str. 13, Bd.  2, 125412 Moscow, Russia}
\eads{\mailto{bogdan.mihalcea@infim.ro}}
\eads{\mailto{dmitrucho@yandex.ru}}

\begin{abstract}
Trapping of microparticles, nanoparticles and aerosols is an issue of major interest for physics and chemistry. We present a setup intended for microparticle trapping in multipole linear Paul trap geometries, operating under Standard Ambient Temperature and Pressure (SATP) conditions. A 16-electrode linear trap geometry has been designed and tested, with an aim to confine a larger number of particles with respect to quadrupole traps and thus enhance the signal to noise ratio, as well as to study microparticle dynamical stability in electrodynamic fields. Experimental tests and numerical simulations suggest that multipole traps are very suited for high precision mass spectrometry measurements in case of different microparticle species or to identify the presence of certain aerosols and polluting agents in the atmosphere. Particle traps represent versatile tools for environment monitoring or for the study of many-body Coulomb systems and dusty plasmas.    
\end{abstract}

\vspace{2pc}
\maketitle
	
\section{Introduction. Particle traps as tools for complex, one-component plasmas (OCP)}

Complex (dusty) plasmas represent a distinct type of low-temperature plasmas that consist of highly charged nano- or microparticles \cite{Fortov2007, Tsyto2008,  Bonitz2014, Campa2014}. Dusty plasmas are encountered in interstellar space and circumstellar clouds, as interplanetary dust or even in the earth magnetosphere, atmosphere and mezosphere \cite{Shukla2002, Fortov2010, Fortov2011, Bonitz2014, Popel2011}. Collective processes occur in complex plasmas owing to the long range Coulomb interaction between particles characterized by large electrical charges, which leads to the occurrence of strong coupling phenomena in the system \cite{Tsyto2008}.
Complex plasmas are intensively investigated in laboratories, as they are expected to shed new light on issues regarding fundamental physics such as phase transitions, self-organization, study of classical and quantum chaos, pattern formation and scaling \cite{Morfill2009, Bonitz2010a, Fortov2010}. Attention paid to the domain has witnessed a spectacular increase after the discovery of plasma crystals \cite{Thomas1994, Schlipf1996, Dubin1999, Tsyto2007} and the detection of spokes in the rings of Saturn by the Voyager 2 mission in 1980 \cite{Shukla2002, Fortov2010}. Present interest is focused on strongly coupled Coulomb systems of finite dimensions \cite{Vladimirov2005}. Particular examples of such systems would be electrons and excitons in quantum dots \cite{Bonitz2014, Chen2007, Werth2009} or laser cooled ions confined in Paul or Penning type traps \cite{Paul1990, Major2005, Quint2014}. 

First experimental observations of ordered structures consisting of charged iron and aluminium microparticles confined in a Paul trap were reported in 1959 \cite{Wuerker1959}. Phase transitions occurred, as an outcome of the dynamical equilibrium between the trapping potential and the inter-particle Coulomb repulsion. In 1991, an experiment reported the storage of macroscopic dust particles (anthracene) in a Paul trap, operating in air \cite{Winter1991}. 
Electrodynamic traps and ion trapping techniques combined with laser cooling mechanisms \cite{Paul1990, Ghosh1995, Werth2009, Haroche2013} allow scientists to investigate the dynamics of small quantum systems and prepare them in well-controlled quantum states \cite{Chen2007, Quint2014}. 
Trapped ions or particles represent one-component plasmas (OCP). The OCP model is a reference one for the study of strongly coupled Coulomb systems \cite{Dubin1999, Davidson2001, Tamashiro1999, Ott2014}. Ion traps have also opened new horizons towards performing investigations on the physics of few-body phase transitions \cite{Schlipf1996, Shukla2002, Fortov2007, Tsyto2008, Bonitz2010a}. Quantum engineering has opened new perpectives in quantum optics and quantum metrology \cite{Zagoskin2011, Haroche2013}. Applications of ion traps span mass spectrometry, very high precision spectroscopy, quantum physics tests, study of non-neutral plasmas \cite{Davidson2001, Dubin1999}, quantum information processing (QIP) and quantum metrology \cite{Kim2005, Chen2007, Leibf2007, Haroche2013, Quint2014}, use of optical transitions in highly charged ions for detection of variations in the fine structure constant \cite{Werth2009, Quint2014} or very accurate optical atomic clocks \cite{Riehle2004, Major2007, March2010, Poli2013, Ludlow2015}. 

In case of quadrupole traps, the second-order Doppler effect is the result of space-charge Coulomb repulsion forces acting between trapped ions of like electrical charges. The Coulombian forces are balanced by the ponderomotive forces produced by ion motion in a highly non-uniform electric field. For large ion clouds most of the motional energy is found in the micromotion. Multipole ion trap geometries significantly reduce all ion number-dependent effects resulting through the second-order Doppler shift, as ions are weakly bound with confining fields that are effectively zero through the inner trap region and grow rapidly near the trap electrode walls. Owing to the specific map shape of trapping fields, charged ions (particles) spend relatively little time in the high RF electric fields area. Hence the RF heating phenomenon (micromotion) is sensibly reduced. Multipole traps have also been used as tools in analytical chemistry to confine trapped molecular species that exhibit many degrees of freedom \cite{Trippel2006, Gerlich2008a, Gerlich2008b, Wester2009}. Space-charge effects are not negligible for such traps. Nevertheless, they represent extremely versatile tools to investigate the properties and dynamics of molecular ions or to simulate the properties of cold plasmas, such as astrophysical plasmas or the Earth atmosphere. 

Stable confinement of a single ion in the radio-frequency (RF) field of a Paul trap is well known, as the Mathieu equations of motion can be analytically solved \cite{Ghosh1995, March2005}. This is no longer the case for high-order multipole fields, where the equations of motion do not admit an analytical solution. In such case particle dynamics is quite complex, as it is described by non-linear, coupled, non-autonomous equations of motion. The solutions for such system can only be found by performing numerical integration \cite{Riehle2004, Major2005, Fortov2010}.

The paper introduces a 16 electrode trap geometry, designed with an aim to levitate and study microscopic particles, aerosols and other constituents or polluting agents encountered in the atmosphere. The  paper is structured as follows: Section \ref{aerosols} introduces general considerations on the importance of dedicated studies on aerosols and nanoparticles in connection with their impact on the global climate, environment protection and human health. Section \ref{spectro} describes the use of ion traps in mass spectroscopy. In Section \ref{exp} we present the trap geometry setup consisting of 16 electrodes.  Maps of the potential within the trap for different supply voltages are given in Section \ref{map}. Section \ref{model} approaches physical modeling and results of numerical simulations performed. Finally, Section \ref{concl} concludes the results of the investigations.

\section{Aerosols. Micro and nanoparticles. General considerations}\label{aerosols}
\subsection{Aerosols and nanoparticles. Global climate and environment protection}

Atmospheric aerosols or astrophysical dusty plasmas are made out of microparticles \cite{Seinfeld2006}. Aerosols are presumed to have a larger impact on climate compared to greenhouse gases, according to the IPCC WGI 4-th Assessment Report \cite{IPCC}. Nevertheless there still is a high uncertainty about it, owing to the aerosol complex composition and a still incomplete picture needed to characterize the interactions between aerosols and global climate. We distinguish between two types of interaction: direct and indirect interactions. By indirect effect, hydrophilic aerosols act as cloud condensation nuclei (CCN) affecting cloud cover and implicitly the radiation balance. Direct interactions account for the light scattering mechanism on aerosols, resulting in cooling effects. On the other hand, aerosols containing black carbon (BC) or other substances absorb incoming light thus heating the atmosphere. According to measurements, the direct radiative effect of BC would be the second-most important contributor to global warming, after absorption by CO$_2$. There is a large interest towards minimizing the uncertainties associated with data collection when evaluating the impact of aerosols on global climate \cite{Kirch2008, Davidovits2008, Lebedev2012}.  Different approaches and methods have been developed for analyzing particles ranging from 10 nm to 10 μm in diameter size, which consist of salts, soot, crustal matter, metals, and organic molecules, often mixed together \cite{Nash2006, Wang2006}. 

Investigations on atmospheric aerosols, viruses, bacteria, and chemical agents, can be performed using high precision mass measurements for micro and nanoparticles \cite{Pandis1995, Seinfeld2006, Kulkarni2011}. Study of such mesoscopic systems is of large interest, as mesoscopic physics is linked to the fields of nanofabrication and nanotechnology. Late research indicates that nanoparticules are also associated with toxic effects on humans \cite{IRSST}, as they are widely used by the cosmetics industry. Many sunscreens contain nanoparticles of zinc oxide or titanium dioxide. There are manufacturers that have added C$_60$ fullerenes into anti-aging creams, as these particles can act as antioxidants. Strong evidence suggests that normally inert materials can become toxic and damaging when they are nano-sized. Evidence collected indicates that nanoparticle effects on human (living) tissue are extremely dangerous \cite{IRSST}. Special care must be taken for people occupied in this area or exposed to nanomaterials, which strongly motivates the need to further analyze and characterize nanomaterial systems \cite{Seo2003, Wang2006}.
  
\subsection{Fine particulate matter and coarse particle pollution}

Microparticles are known to inflict harmful effects on humans. There is a strong effort towards limiting the maximum concentration and enforce safety levels for the atmospheric microparticles or dust most hazardous to human health. We distinguish between two categories: (i) fine particles with a diameter less than 2.5 microns (also called fine particulate matter or PM$_{2.5}$), which are the most dangerous, and (ii) larger particles with a diameter less than 10 microns but larger than 2.5 microns, namely the particulate matter PM$_{10}$ (also called coarse particles). Important steps are being taken to reduce pollution due to PM$_{2.5}$ and PM$_{10}$ microparticles, in order to minimise their harmful effects on humans and biological tissue. In the EU directives are enforced which regulate the PM$_{2.5}$ and PM$_{10}$ microparticle levels. Member States must set up {\em sampling points} in urban and also in rural areas. Besides particulate matter, these sampling points must perform measurements on the concentration of sulphur dioxide, nitrogen dioxide and oxides of nitrogen, lead, benzene and carbon monoxide. 

Strong evidence indicates that breathing in PM$_{2.5}$ over the course of hours to days (short-term exposure) and months to years (long-term exposure) can cause serious public health effects that include premature death, adverse cardiovascular and respiratory effects or even harmful developmental and reproductive effects \cite{EPA2012}. Lung cancer is associated with the emission of microparticles produced by diesel engines \cite{Kirch2008}. Scientific data also indicates that breathing in larger sizes of particulate matter (coarse particles or PM$_{10}$), may also have public health consequences. In addition, particle pollution degrades public welfare by producing haze in cities or constantly increasing the rate of allergies for population living in urban areas. People with obesity or diabetes are more vulnerable to increased risk of PM - related health effects. This is why constant monitoring of the various polluting agents is a major concern for health services and life quality assurance in Europe and in other countries throughout the world \cite{IRSST, EPA2012}.

\section{Mass spectrometry using ion traps}\label{spectro}

A linear Paul trap uses a superposition of time varying, strongly inhomogeneous (a.c.) and d.c. electric potentials, to achieve a trapping field that dynamically confines ions and other electrically charged particles \cite{Gerlich1992, March2005, Major2005, Chen2007, Haroche2013}. When the a.c. trapping voltage frequency lies in the few Hz up to MHz or even GHz range, electrons, molecular ions or electrically charged nanoparticles with masses of more than $10 u$ (atomic mass units) are confined\cite{Gerlich2003, Gerlich2008b, Wester2009, March2010}. Ion dynamics in Paul traps is described by a system of linear, uncoupled equations of motion (Hill equations) that can be solved analytically \cite{Ghosh1995, March2005}. The linear form of the trap can be used as a selective mass filter or as an actual trap by creating a potential well for the ions along the $z$ axis of the electrodes \cite{Gerlich1992, Gerlich2008a, Otto2009}. The linear trap design results in increased ion storage capacity, faster scan times, and simplicity of construction \cite{Trevi07, Nie2008}. A Paul trap runs in the mass-selective axial instability mode by scanning the frequency of the applied a.c. field \cite{Nie2008, Trevi09}. Microparticle diagnosis can be achieved by operating the Quadrupole Ion Trap (QIT) as an electrodynamic balance, for low frequency values of the a.c. field applied to the trap electrodes (typically less than 1 kHz).  

A specific charge $m/z$ can be isolated in the ion trap by ejecting all other $m/z$ particles (ions), by applying various resonant frequencies. Moreover, an ion trap can be coupled to an Aerosol Mass Spectrometer (AMS) to investigate atmospheric aerosol (nano)particles.  
Experimental setups based on multipole Paul traps enable confinement as well as qualitative investigation of atmospheric particles and aerosols \cite{Lebedev2012}. Using such instruments, the chemical composition of the non-refractory component of aerosol particles can be measured quantitatively. While the AMS uses either a linear quadrupole mass filter (Q-AMS) or a time-of-flight mass spectrometer (ToF-AMS) as a mass analyzer, the Ion Trap IT-AMS uses a 3D quadrupole ion trap. The main advantages of an ion trap are the possibility of performing MS-experiments as well as ion/molecule reaction studies \cite{Vogel2013}. Experiments demonstrate that a mass resolving power larger than 1500 can be achieved. This value is high enough to separate different organic species at $m/z ~ 43$. Calibrations with laboratory-generated aerosol particles indicate a linear relationship between signal response and aerosol mass concentration. These studies, together with estimates of the detection limits for particulate sulfate ($0.65 \mu$g/m$^3$) and nitrate ($0.16 \mu$g/m$^3$) demonstrate the ability of the IT-AMS to measure atmospheric aerosol particles \cite{Gerlich1992, Trevi2006, Wester2009, Kurten2007, Smith2008, Trevi09}.   

\section{Experimental Setup}\label{exp}

\begin{figure}[h!tb]
\centering
\includegraphics[width=0.45\textwidth]{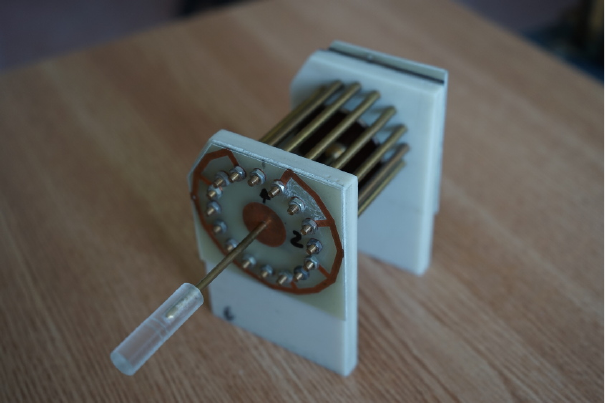}
\caption{Photo of 16-electrode linear Paul trap}
\label{iontraps}
\end{figure}

The 16-electrode trap geometry setup we have tested is shown in Fig. \ref{iontraps}. The trap geometry consists of 16 brass electrodes of 60 mm length and 4 mm diameter, equidistantly spaced on a 46 mm diameter. If the AC potential is not too large ($ V_{ac} > 3.5$ kV), stable oscillations will occur until the d.c. potential is adjusted to balance vertical forces such as gravity. When such condition is fulfilled, the oscillation amplitude becomes vanishingly small and the particle is stable trapped \cite{Seinfeld2006}. Electrically charged microparticles are trapped in air, at Standard Ambient Temperature and Pressure (SATP) reference conditions, which results in an efficient {\em cooling} of the particle owing to friction in air \cite{Winter1991, Izmailov1995}. Such mechanism is similar with cooling of ions in ultrahigh vacuum conditions by means of collisions with the buffer gas molecules. The 16-electrode trap geometry we report is intended for investigating complex Coulomb systems (microplasmas), such as microscopic particles or aerosols in the atmosphere. We bring new evidence on microparticle trapping, while demonstrating that the stability region for multipole traps is larger with respect to a quadrupole trap (electrodynamic balance configuration) \cite{Vasilyak2013, Lapitsky2015}. 

A radiofrequency (RF) voltage (typically between $1 \div 2.5$ kV)  generates an oscillating quadrupole potential in the $y-z$ plane, responsible for radial trapping of particles in a 2D potential \cite{Paul1990, Ghosh1995, Werth2009}. A d.c. potential applied between the two {\em endcap electrodes} located along the trap $x$ axis, helps achieving axial confinement for positively charged particles. A harmonic secular potential results within the trap if both the RF and d.c. endcap potentials are quadratic, a condition difficult to achieve for the whole trap volume. In fact, it is assumed that the potential in the vicinity of the trap axis is harmonic, which is a sufficiently accurate approximation. The 16-electrode Paul trap geometry is intended for investigating the appearance of stable and ordered patterns for different charged microparticle species. Preliminary tests are performed using alumina microparticles (with dimensions ranging from 60 up to 200 microns) in order to illustrate the trapping phenomenon, but other species can be considered. Specific charge measurements for trapped microparticle and nanoparticle species are expected to result, as it is our intention to refine the setup. The trap we have designed is 60 mm long, its 16 bars are equidistantly spaced within a 46 mm diameter. The 16-electrode trap we have designed exhibits a variable geometry as shown in Fig. \ref{iontraps}.  

The supply system currently under test consists of two independent units: one of them supplies the a.c. trapping voltage applied between the trap rods. The a.c. voltage  achieves radial trapping of the charged particles. It is applied between the even and uneven numbered trap electrodes. The a.c. voltage is generated using a step-up transformer able to generate 3.5 kV. Typical trapping voltages lie in the $1.5 \div 2.5$ kV range. A low frequency oscillator will be used to drive the step-up transformer, at a frequency value that can vary between 45 Hz up to few hundred Hz.
The second unit of the electronic supply system consists of a d.c.-d.c. converter delivering a variable d.c voltage $U_x$ (0-1000 V), applied between the upper and lower multipole trap electrodes. The $U_x$ voltage (also called diagnose voltage) compensates the particle shift owing to the gravitational field and also enables performing a diagnose of the trapped microparticles. By using a precision microscope to ascertain the $x$-axis shift of the microparticle as a function of the $U_x$ voltage variation, the specific charge of the microparticle species can be determined. Ions confined in Paul traps that operate in ultrahigh vacuum arrange themselves along the longitudinal $z$ axis as they are contained  within a large region located around it, where the trapping potential is very weak. Things are different in case of trapped microparticles, which we explain in Section \ref{concl}. Another d.c. variable voltage $U_z$ is applied between the trap endcap electrodes, in order to achieve axial confinement and prevent particle loss near the trap ends. The polarity of the d.c. voltages can be reversed. Both a.c. and d.c. units of the electronic supply system will be driven by a microcontroller unit.

\section{Trap potential map for the 16 pole trap}\label{map}

\begin{figure}[h!tb]
\centering
\includegraphics[width=0.6\linewidth]{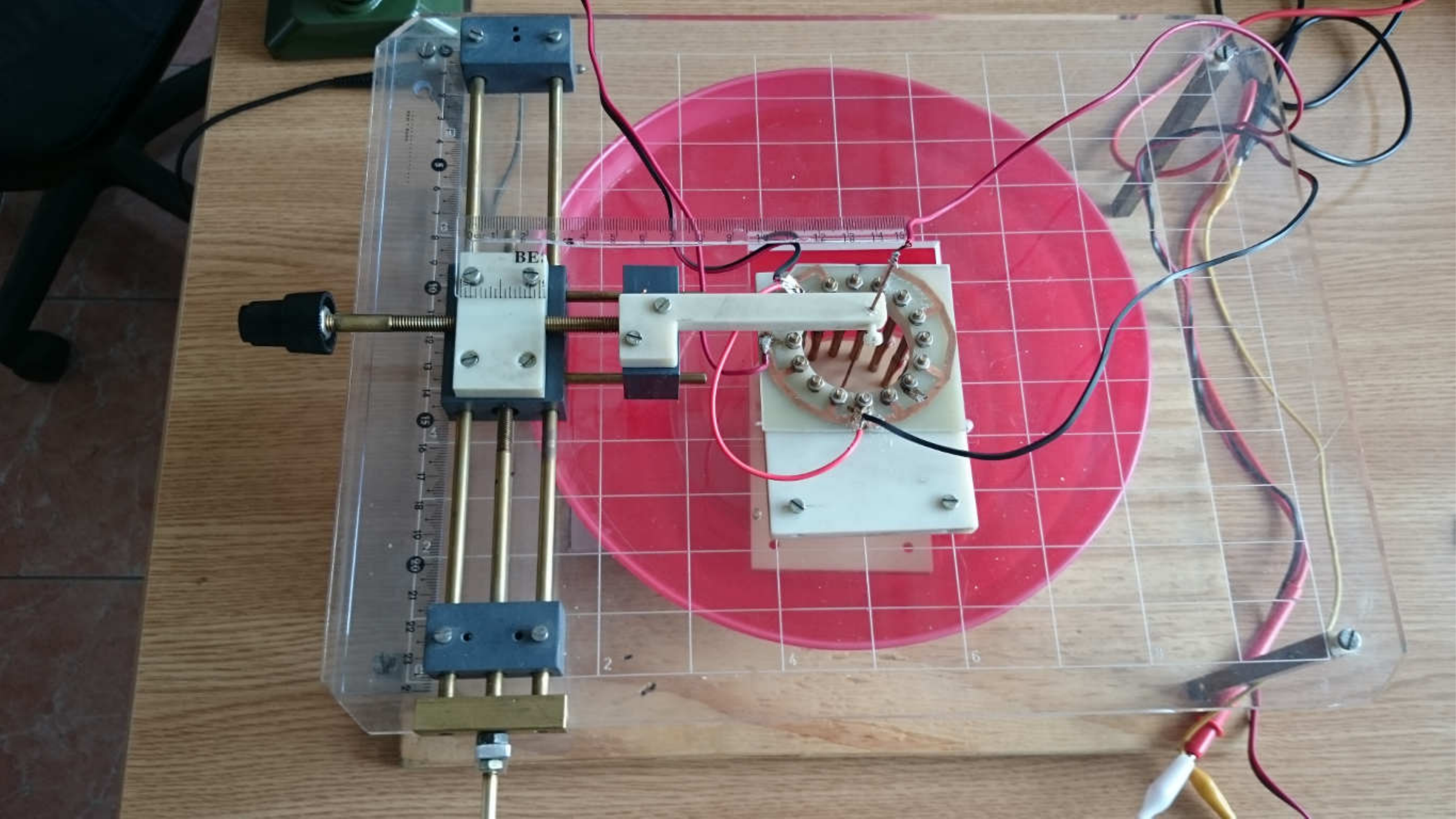}
\caption{Experimental setup photo showing the 16-electrode trap, the electrolytic tank and the precision mechanism used to chart the electric potential}
\label{expsetup}
\end{figure}

 The a.c. potential within the trap was mapped using an electrolytic tank filled with dedurized water. The trap was immersed within the water around 38 mm of its total length. A needle electrode was used to measure the trap voltage, immersed at 16 mm below water level. The needle electrode can be displaced 36 mm both horizontally and vertically, using a precision mechanism. The trap potential was mapped using a transversal (radial) section located at 22 mm with respect to the immersed end, for 2 mm steps on both vertical and horizontal position. The experimental setup including the trap and the electrolytic tank is shown in Fig. \ref{expsetup}.

The Paul trap was supplied with a sine wave delivered by a function generator, at a 46.1 Hz frequency. An oscilloscope was used to monitor the sine wave. The rms value of the a.c. voltage was measured using a precision voltmeter. We supply below maps of the trap potential for a 0.5 V amplitude sine wave supplied to the trap electrodes. The sine wave was applied between even and respectively uneven electrodes, connected together. 

\begin{figure}[h!tb]
\centering
\includegraphics[width=0.45\linewidth]{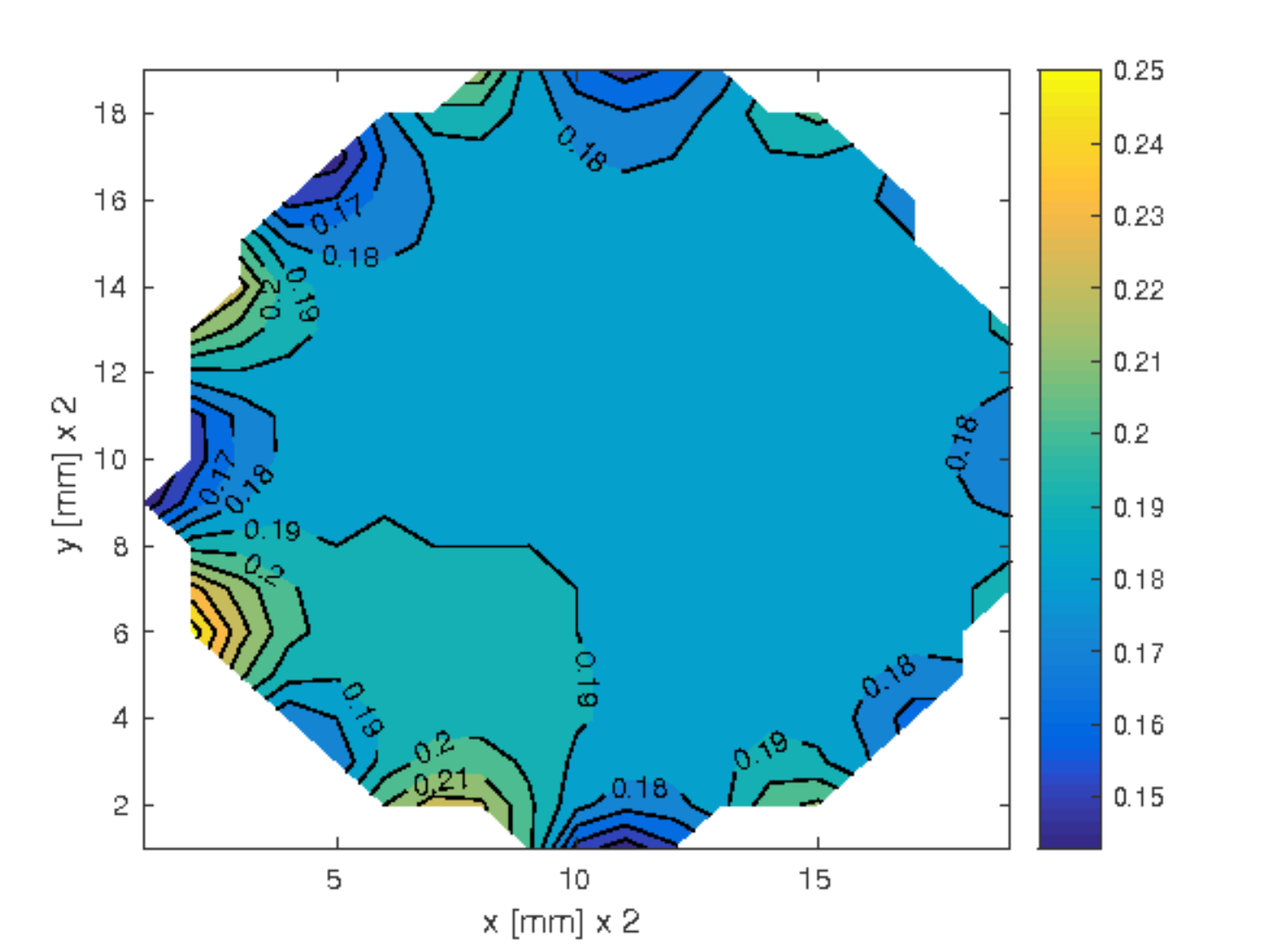}
\includegraphics[width=0.45\textwidth]{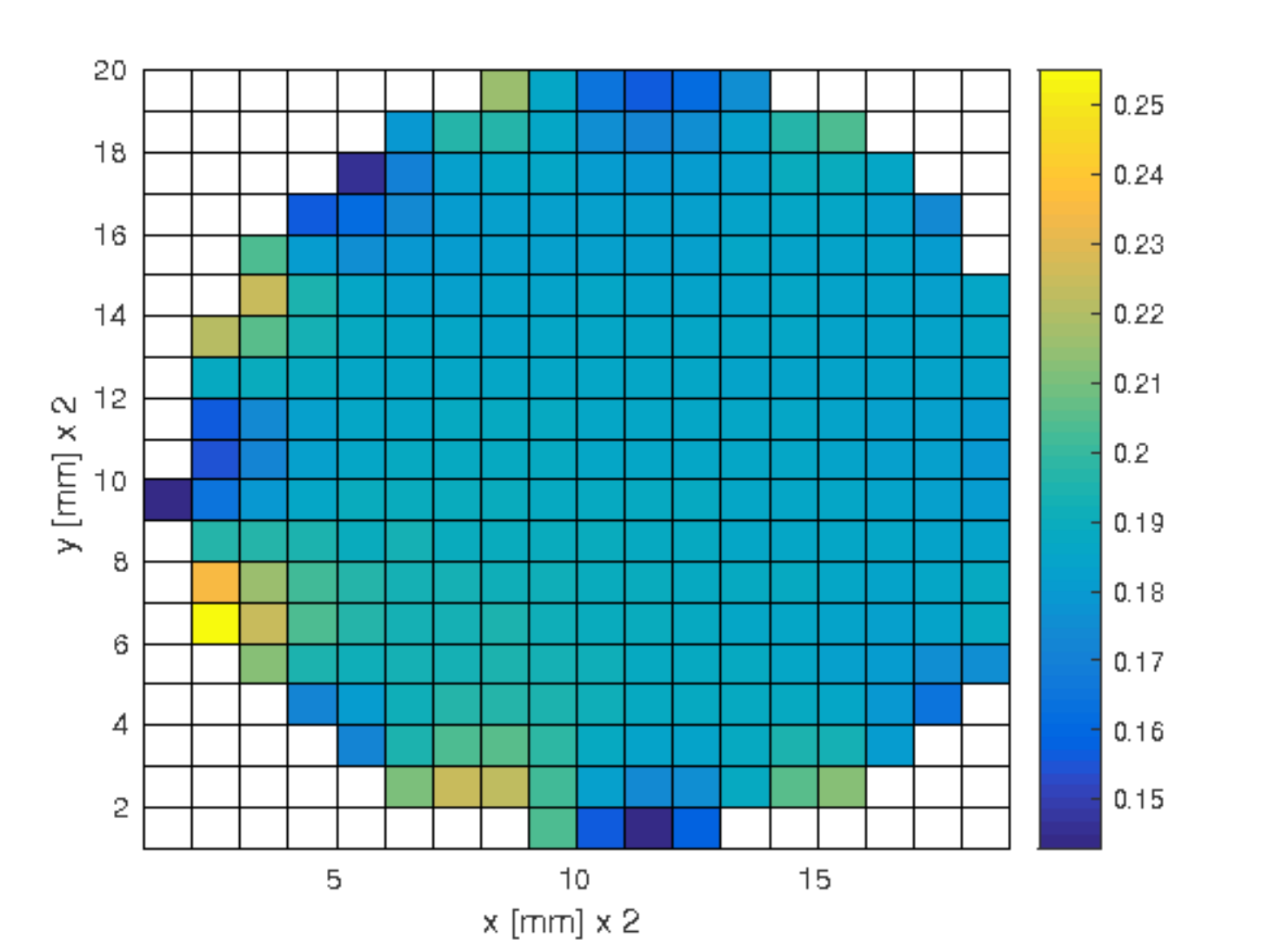}
\includegraphics[width=0.47\textwidth]{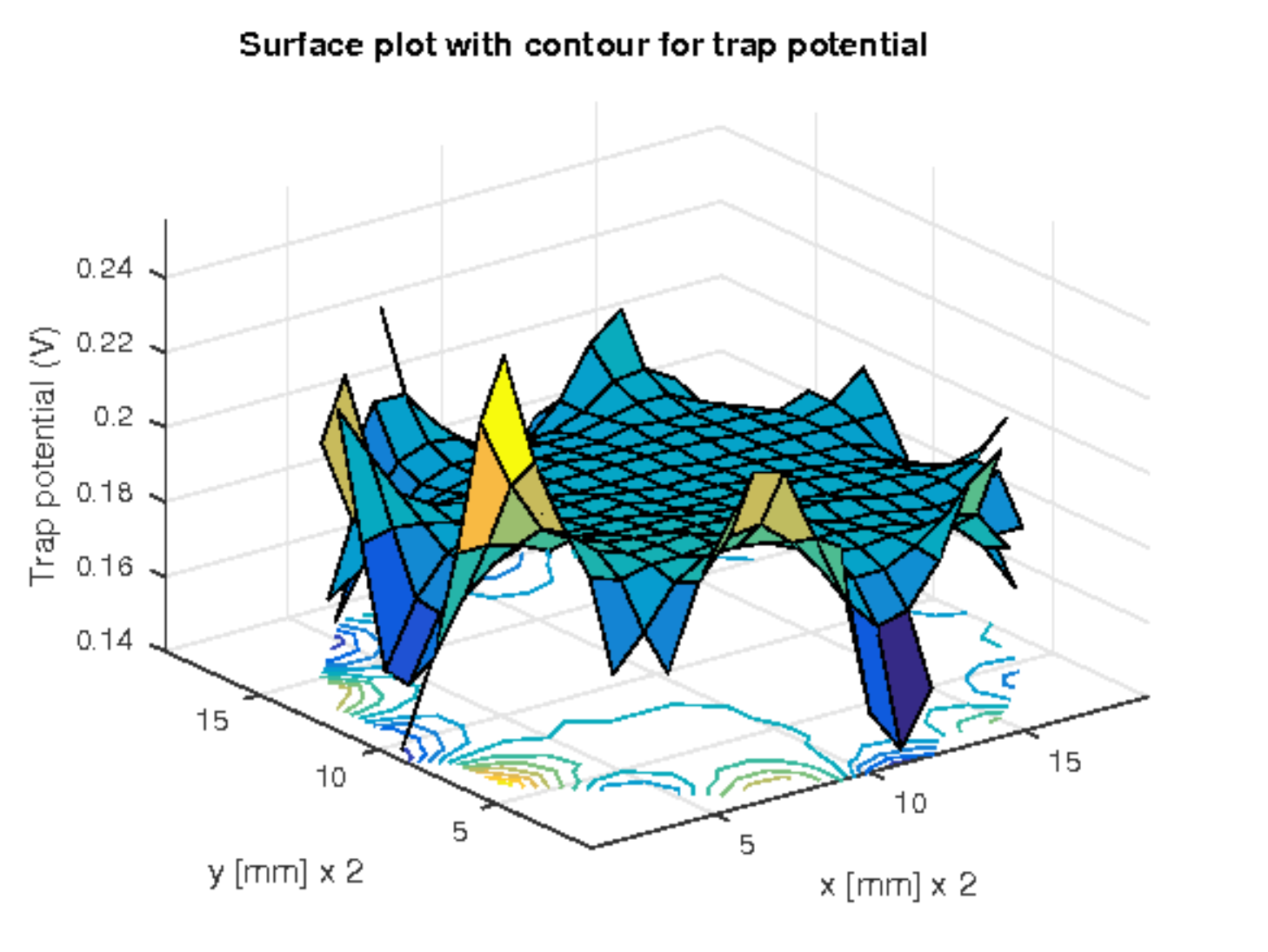}
\includegraphics[width=0.47\textwidth]{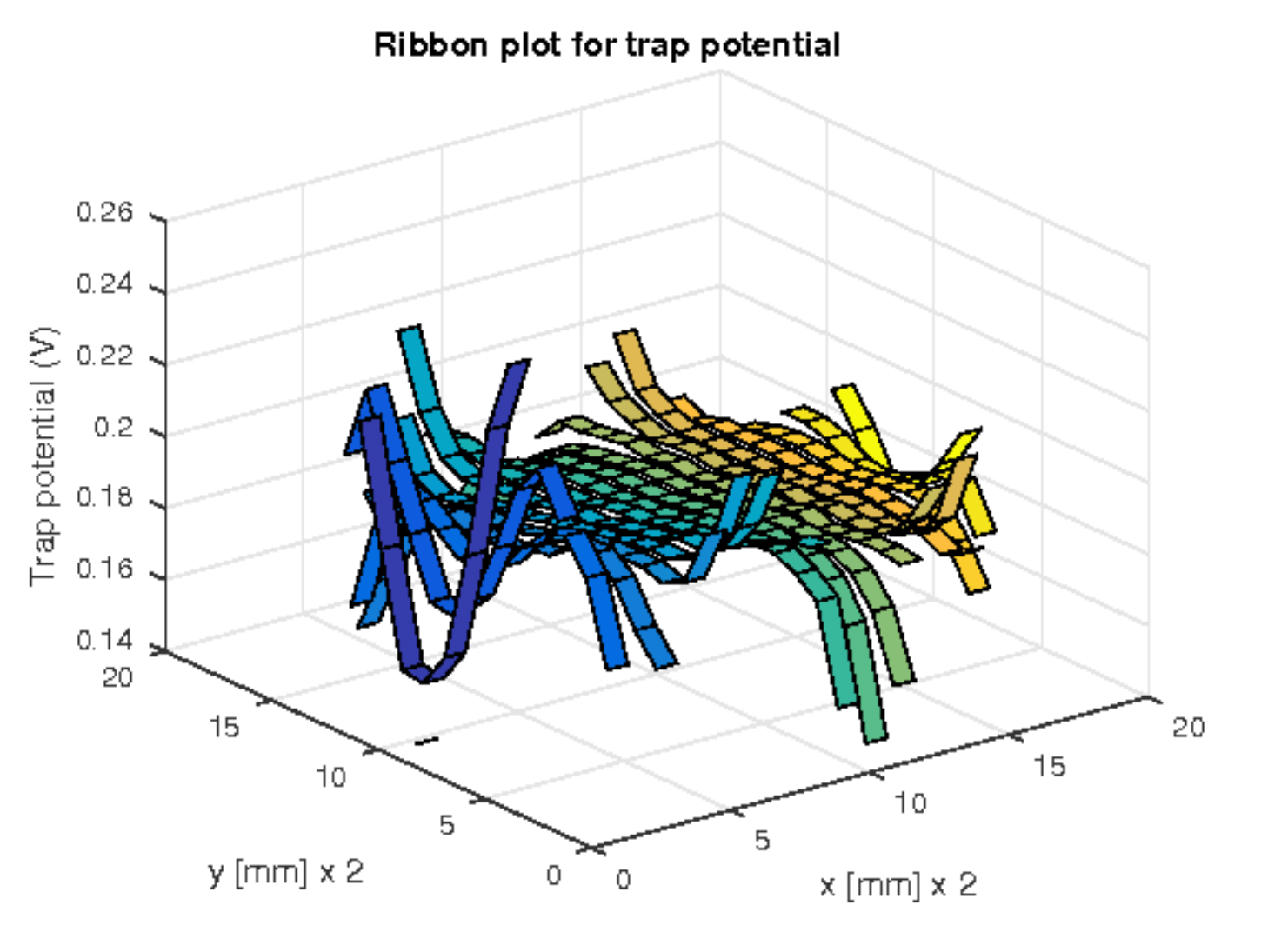}
\caption{Maps of the 16-electrode trap potential: contour plot, pseudocolor plot, surface plot and ribbon plot for an input sine wave of 0.5 V amplitude (0.335 V rms)}
\label{mapchart05V}
\end{figure}

We have also mapped the trap potential for a supply voltage of 1.5 V amplitude (1.037 V rms value) sine wave. 

\begin{figure}[h!tb]
\centering
\includegraphics[width=0.45\linewidth]{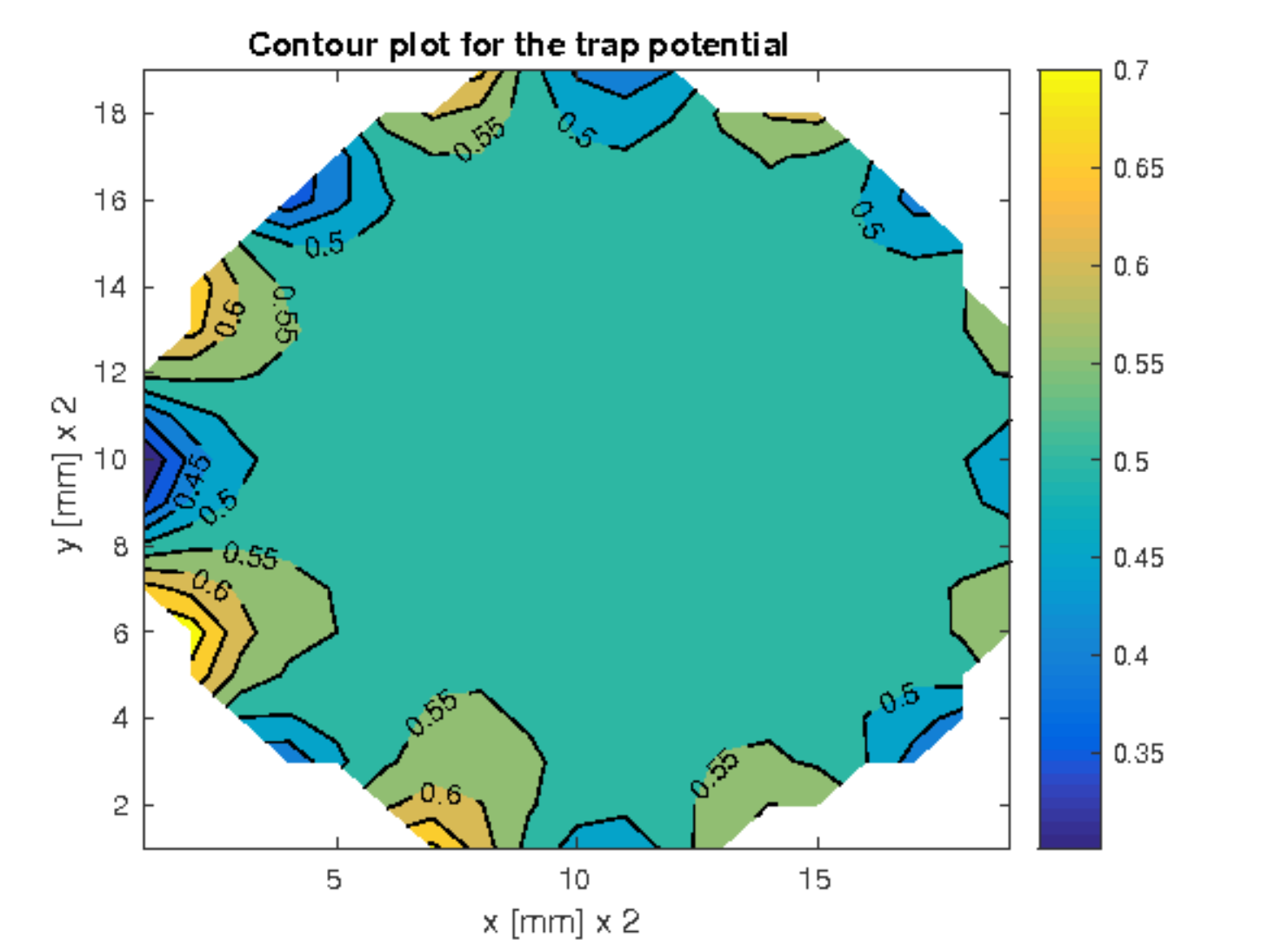}
\includegraphics[width=0.5\textwidth]{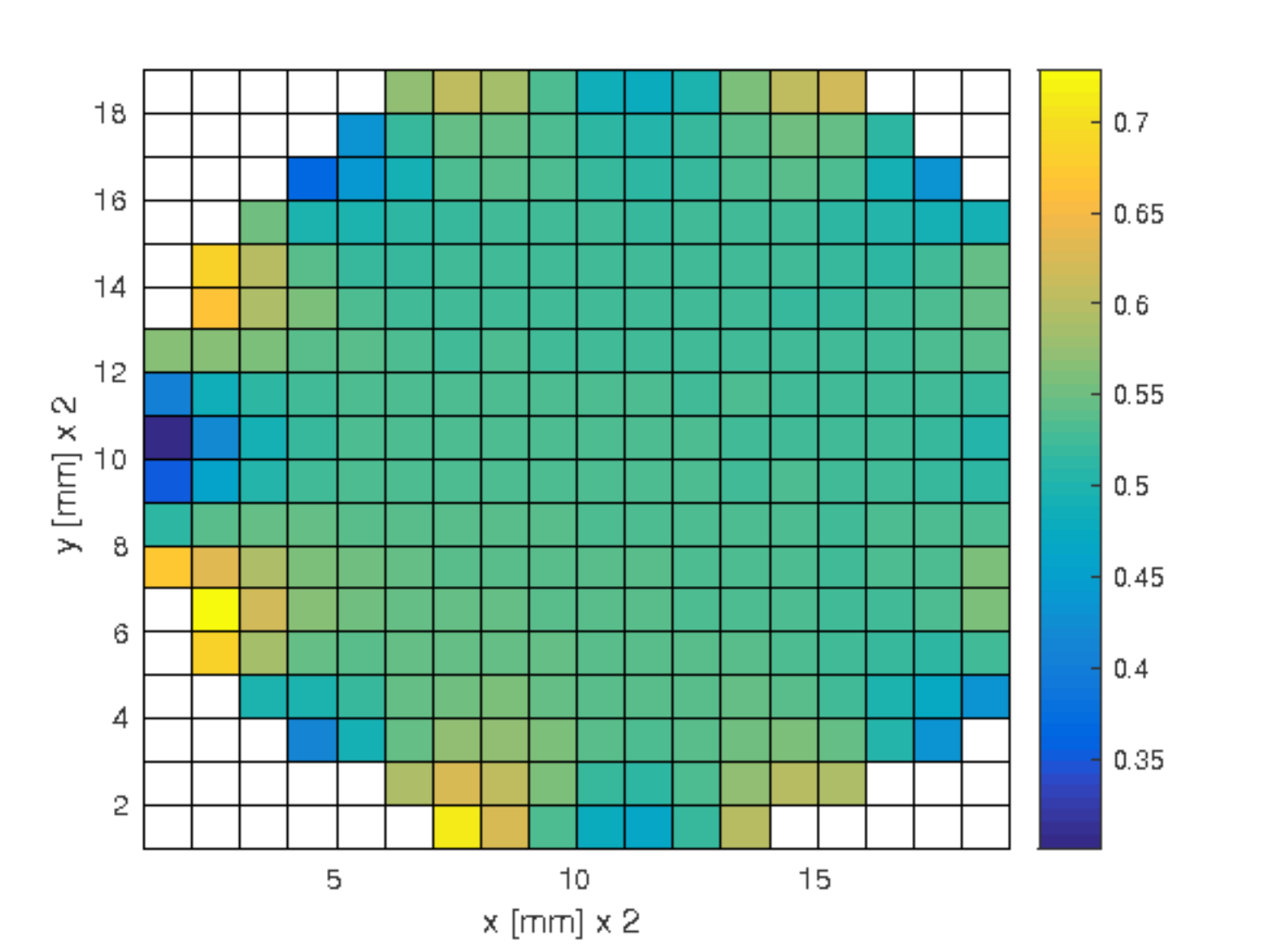}
\includegraphics[width=0.48\textwidth]{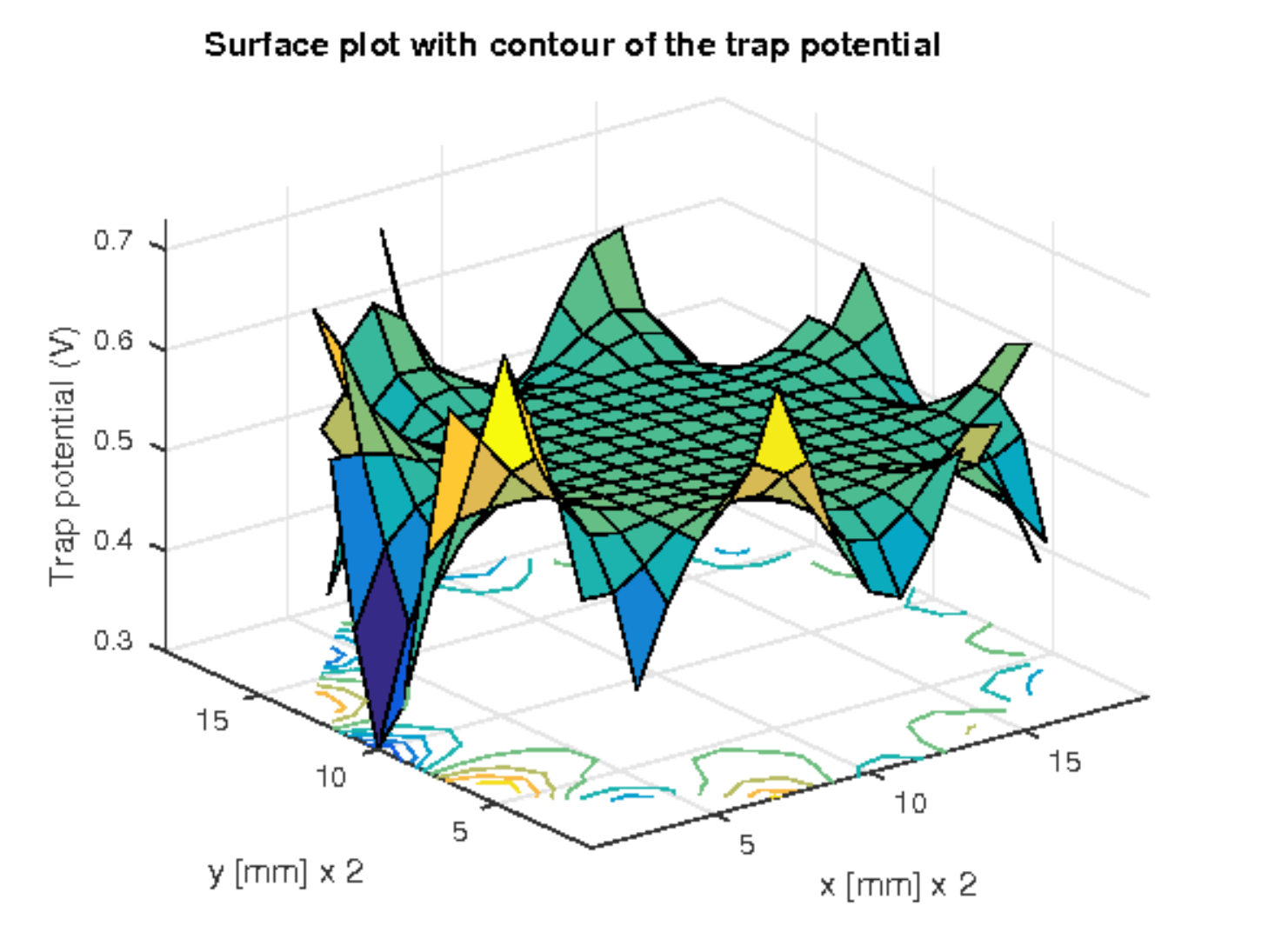}
\includegraphics[width=0.48\textwidth]{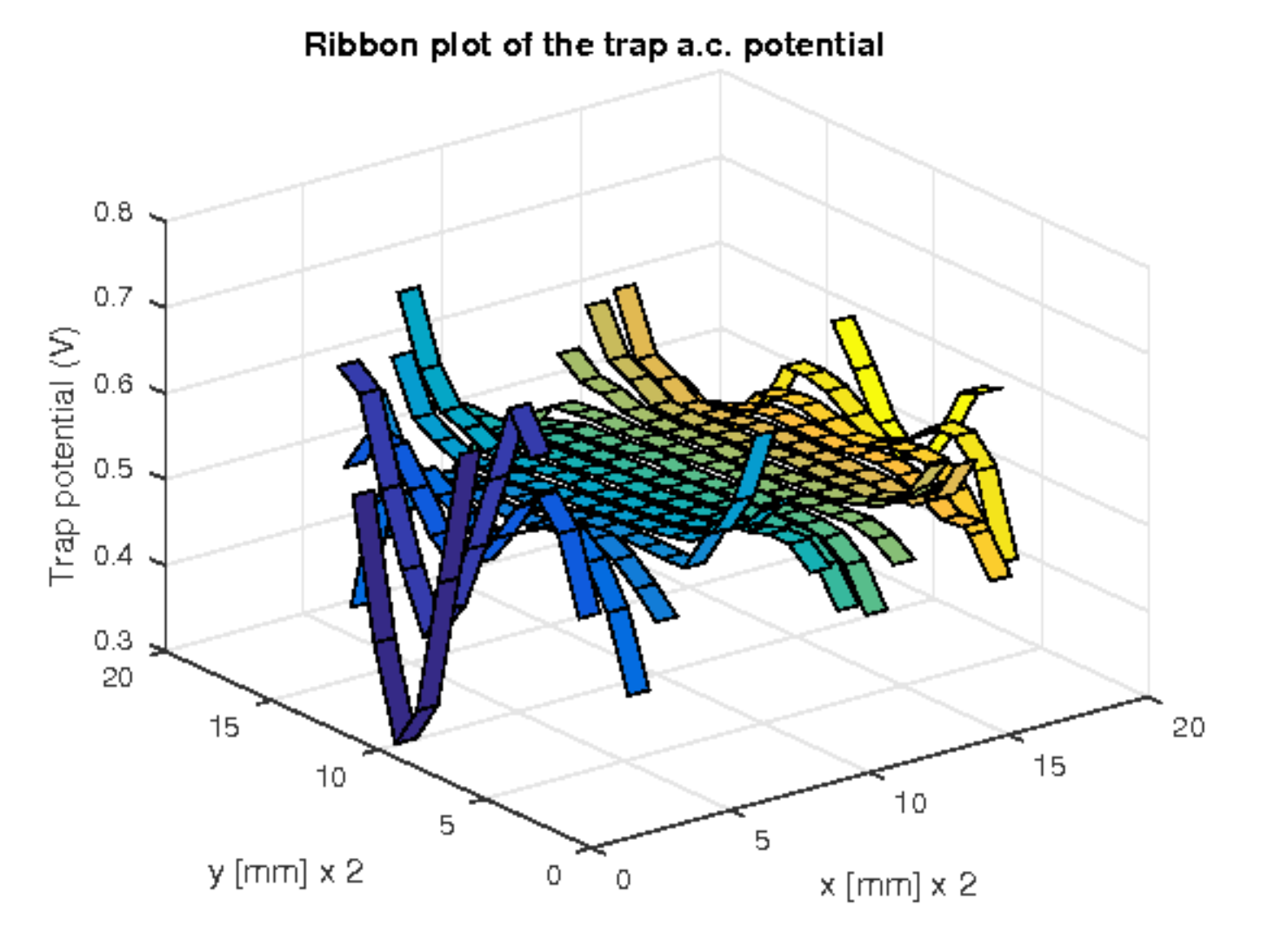}
\caption{Maps of the 16-electrode trap potential: contour plot, pseudocolor plot, surface plot and ribbon plot for an input sine wave of 1.5 V amplitude (1.037 V rms)}
\label{mapchart15V}
\end{figure}

Data presented is preliminary, as it is our goal to demonstrate that such trap design is characterized by an extended region of lower field compared with 8-electrode and 12-electrode geometries we have tested. Moreover, the 16-electrode trap design is suitable for various applications where larger signal-to-noise ratios are required. 
The experimental data and the numerical simulation results suggest that linear particle traps, such as the ones we have investigated \cite{Gheorghe98, Stoican2008, Mihalcea2008, Visan2013, Vasilyak2013, Lapitskiy2015}, are suited to use as Ion Trap Aerosol Mass Spectrometers (IT-AMS) \cite{Trevi2006, Vogel2013}. Alumina (Al$_2$O$_3$) microparticles have been confined in our experiments using three dimensional electrodynamic multipole fields. The multipole trap geometries we report, have been investigated with an aim to levitate and study microscopic particles, aerosols and other constituents or polluting agents that are present in the atmosphere. The research and simulations performed are based on previous results and experience \cite{Gheorghe98, Stoican2008, Visan2013, Vasilyak2013, Lapitskiy2015}.

\section{Physical modelling and computer simulation}\label{model}

Besides experimental investigations, we have performed numerical simulations in order to illustrate particle dynamics in multipole traps. Our main concern was to choose conditions as close as possible to the experiment, in an attempt to validate experimental data. Our simulations consider stochastic forces due to random collisions with neutral particles, viscosity of the gas medium, regular forces produced by the a.c. trapping voltage and the gravitational force. Thus, microparticle dynamics is described by a stochastic Langevin differential equation, as follows \cite{Shukla2002, Guan2011, Vasilyak2013, We2}:

\begin{equation}
\label{eq.1}
m_p \frac{d^2 r}{dt^2} = F_t(r)-6\pi\eta r_p \frac{dr}{dt} + F_b + F_g
\end{equation}
where $m$ and $r_p$ represent the microparticle mass and radius vector, $\eta$ is the dynamic viscosity of the gas medium with $\eta = 18.2\mu$Pa$\cdot$s, and $F_t(r)$ is the ponderomotive trapping force. The $F_b$ term stands for the stochastic delta-correlated forces accounting for stochastic collisions with neutral particles, while $F_g$ is the gravitational force. We have considered a microparticle with mass density $\rho_p = 3700$ kg/cm$^3$. The numerical method developed in \cite{Skeel2002} was used in order to solve the stochastic differential equation~(\ref{eq.1}). 

The Coulomb force acting on the microparticle (expressed as the sum of contributions for each trap electrode) is the vector sum of forces of point-like charges uniformly distributed along the electrodes, as shown in \cite{Vasilyak2013, Lapitskiy2015}:

\begin{equation}
\label{eq.2}
|F_t(r)|=\sum\limits_s\frac{L V q}{2 N \ln{(\frac{R_2}{R_1})}(r_s - r)^2},
\end{equation}
where $L$ is the length of the trap electrodes, $V$ is the trapping voltage: $V_{ac} \sin (\Omega t)$ or $V_{ac} \sin (\Omega t+\pi)$, $q$ is the microparticle charge, $N$ is the number of point-like charges for each trap electrode, $R_2$ and $R_1$ represent the radii of the grounded cylindrical shell surrounding the trap and trap electrode respectively, while $r$ and $r_s$ denote the vectors for microparticle and point-like charge positions respectively. 

For such model, the results of the computations depend on the following relationship between the relevant trap parameters   
$$\Phi_p = \frac{V_{ac}\: q}{2 \ln(\frac{R2}{R1})} .$$ 

The following trap parameters were chosen: length of electrodes $L = 6.5$ cm, $R_2 = 25$ cm, $R_1 = 3$ mm, trap radius $r_t = 4$ cm and $V_{ac} = 2 $kV. According to Eq.~(\ref{eq.2}), the force $F_t(r)$ describing microparticle interaction with the ponderomotive force depends on the electrical charge $q$, the trapping voltage $V_{ac}$ and the geometrical trap parameters. For such model the regions of microparticle confinement are mainly determined by these forces. However, in reality the regions of microparticle stable confinement depend also on the trap volume, number of trapped microparticles,  average interparticle distance and, as a consequence, on repulsive forces of the interparticle interactions defined by the particle charge $q$. To minimize the influence of these physical factors, we used a large value of the a.c. trapping voltage electrode $V_{ac} = 2$ kV. In such case, the captured particles charge will be smaller and results of the simulations will be more or less universal.  

\begin{figure}[bth]
\begin{center}
\includegraphics[width=0.5\linewidth]{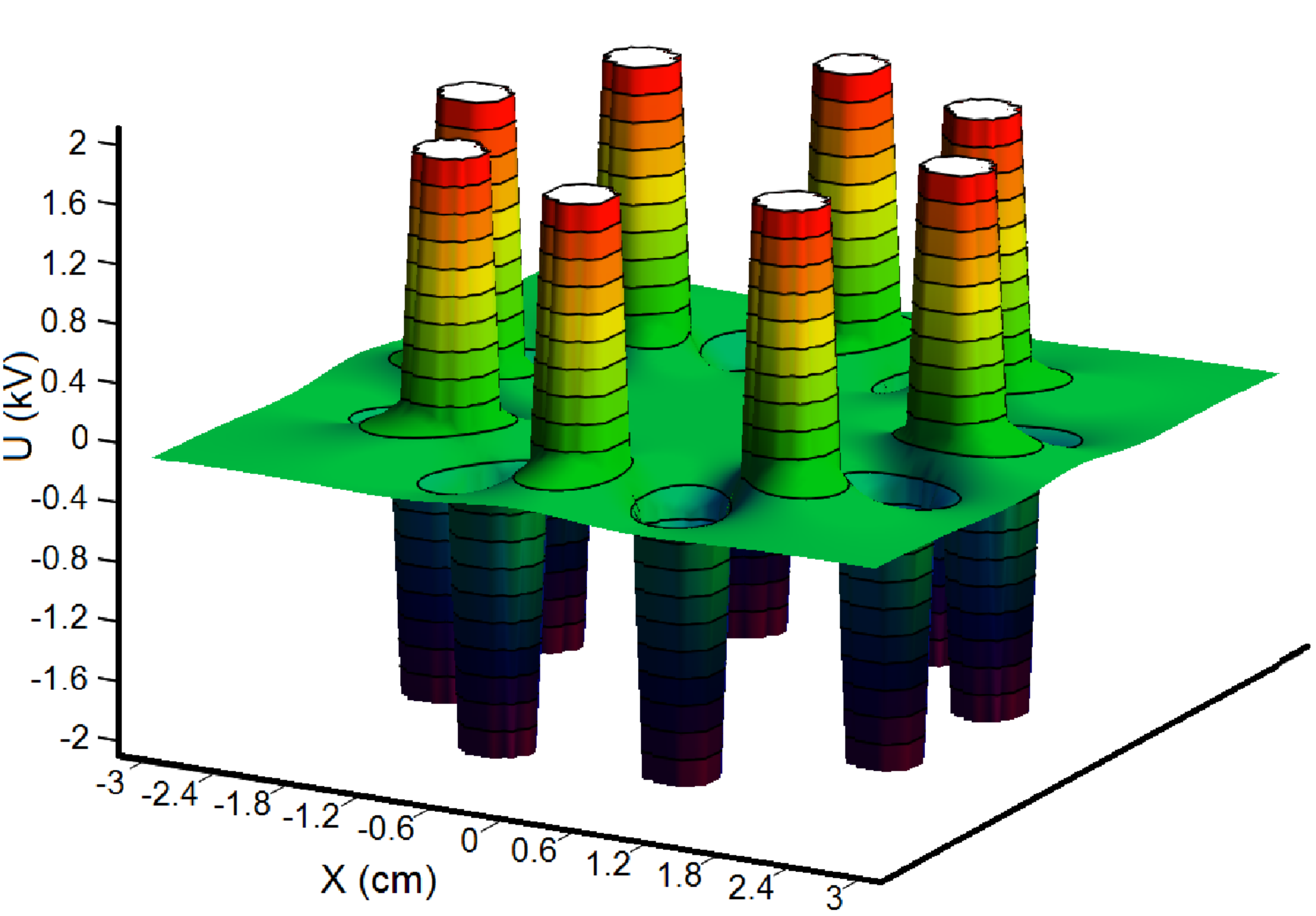}
\end{center}
\caption{3D plots for the 16 electrodes trap.} 
\label{surf}
\end{figure}

The equipotential surfaces in cross section of a linear trap with 16 electrodes is shown in Fig.~\ref{surf}. The equation describing the electric potential can be expressed as:

\begin{equation}
\label{eq.3}
U\left(x,y\right)=\sum\limits_{j=1}^{N_{el}}\sum\limits_{s}\frac{(-1)^jL V_{ac}}{2 N \ln\left(\frac{R_2}{R_1}\right)\sqrt{\left(x - r_t \cos\left(\frac{2\pi j}{N_{el}}\right)\right)^2 + \left(y-r_t \sin\left(\frac{2\pi j}{N_{el}}\right)\right)^2 + {z_s}^2}},
\end{equation}
where $N_{el}$ is the number of trap electrodes and $z_s$ is the $z$ axis coordinate of each point-like charge for the trap electrodes. 

In Fig.~\ref{surf} hills correspond to potential barriers while pits correspond to potential wells. White holes inside the hills correspond to trap electrodes. Every half cycle of the oscillation, barriers and wells swap positions and each charged particle oscillates between them. Thus, particle oscillations result in dynamic confinement \cite{Paul1990}.

\begin{figure}[bth]
\begin{center}  
\includegraphics[width=0.5\linewidth]{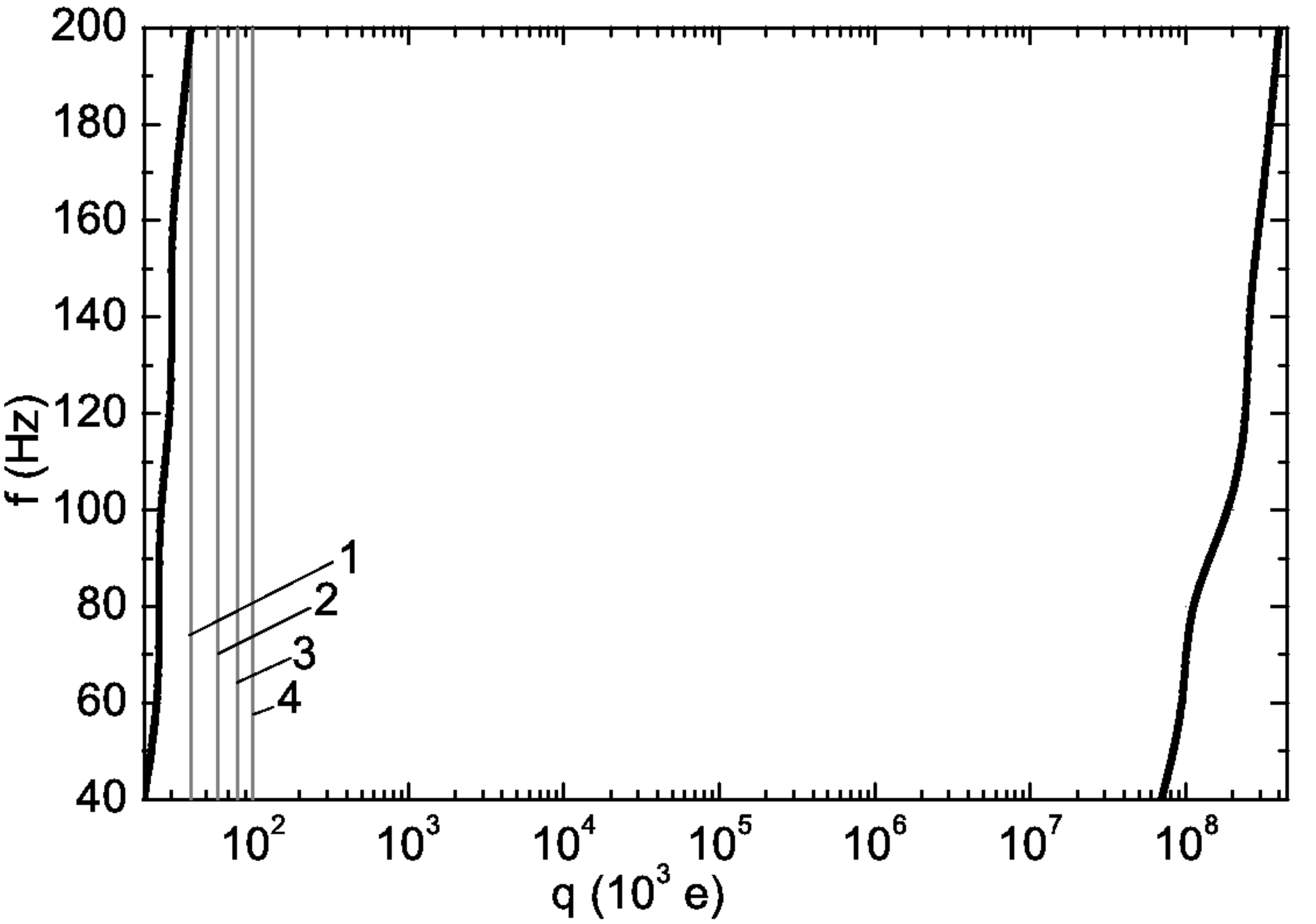}
\end{center}
\caption{The regions of a single particle confinement as a function of the a.c. voltage frequency $f$ and particle charge. Numerical simulations were performed for an average radius $r_p = 5 \mu$m of the microparticle and a particle charge value ranging between $q = 3 \cdot10^4 e$ to $5 \cdot 10^{11} e$. Vertical lines $1 -- 4$ correspond to $q = 4, 6, 8, 10 \cdot 10^4 e$ that have been used to estimate oscillation amplitudes within the regions.}
\label{8-12-16}
\end{figure}

Figure~\ref{8-12-16} presents the confinement region for the charged microparticle. The confinement region is located between black lines. Outside this region, the particle is not trapped. For small particle charge (at the left-hand of the confinement region), the a.c. trapping electric field cannot compensate the gravity force, and particles flow through the trap. At the right-hand of the confinement region when the electric charge value is large, the trap field is strong enough to push microparticles out of the trap.

To study the influence of the frequency and particle charge on the behaviour of particles, we have investigated the average amplitude of particle oscillations within the trap. The dynamics of several microparticles in the trap was studied and averaged amplitudes of particle oscillations have been estimated for a time duration equal with ten periods of the a.c. trapping voltage. 

\begin{figure}[bth]
\begin{center}
\includegraphics[width=0.4\linewidth]{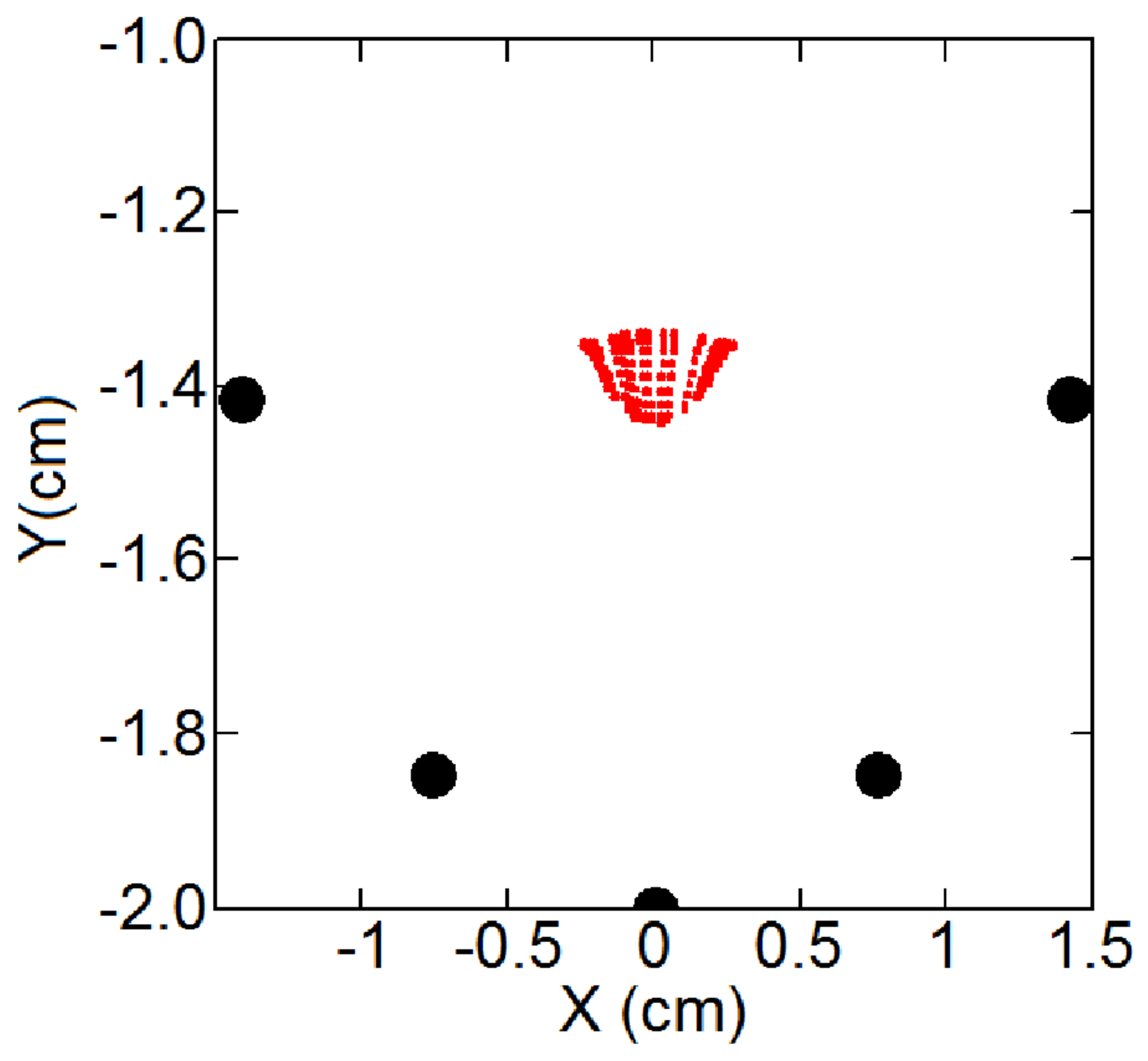}
\end{center}
\caption{End view of the microparticle tracks in a 16-electrode trap at $f = 60$ Hz. Big black dots correspond to trap electrodes (not in scale). The microparticle radius was $r_p = 5 \:\mu$m and the electric charge value $q = 8 \cdot 10^4 e$.}
\label{tracks8-12-16}
\end{figure}

\begin{figure}[bth]
\begin{center}
\includegraphics[width=0.55\linewidth]{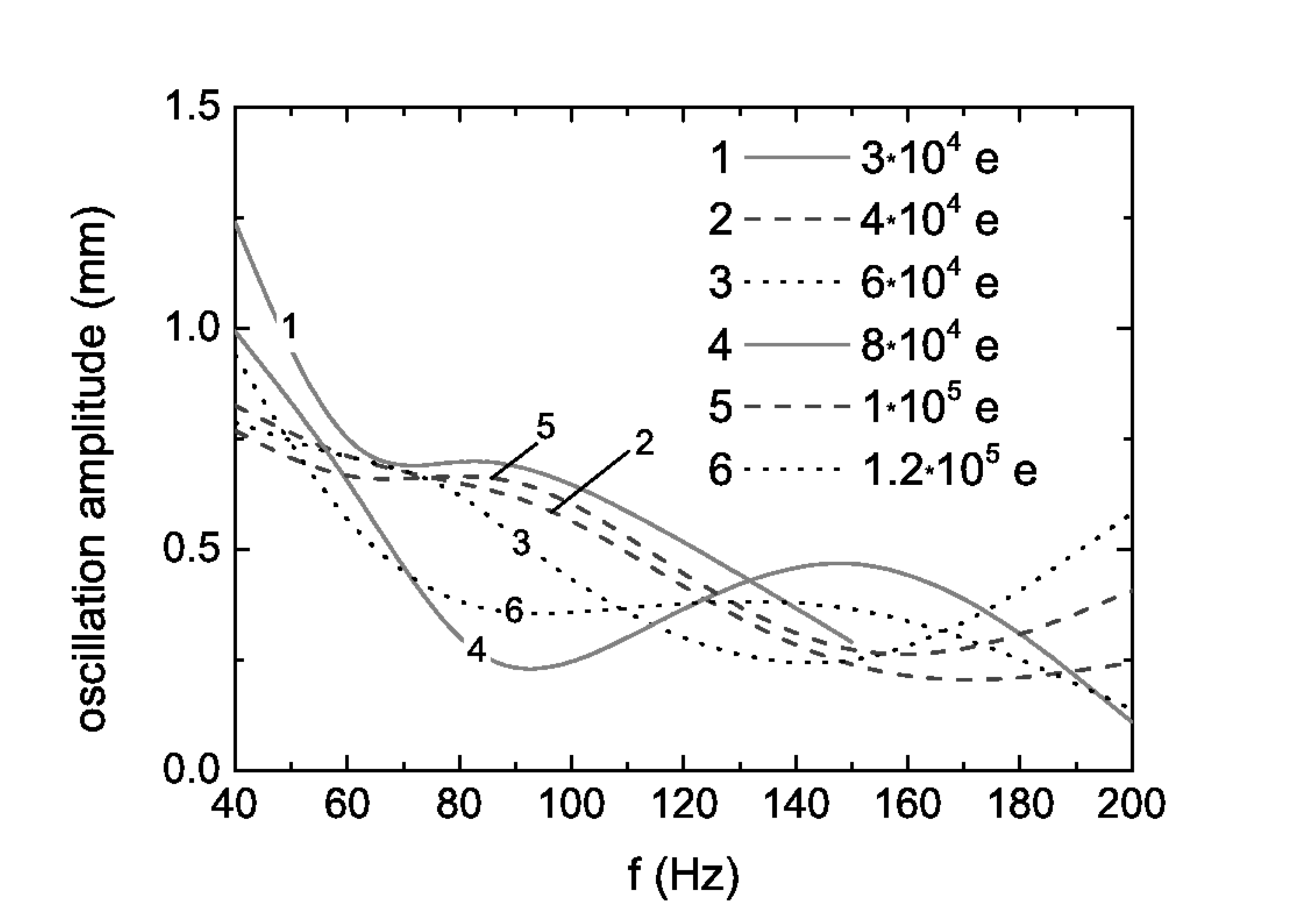}
\end{center}
\caption{Dependencies of the average oscillation amplitude on the frequency and the particle charge $q$. Numerical simulations were performed for parameters used in the experiment such as: particle radius $r_p = 5 \:\mu$m and electric charge value ranging from $q = 3 \cdot 10^4 e$ to $1.2\cdot 10^5 e$.}
\label{ampl1}
\end{figure}

Particle oscillations are presented in Fig.~\ref{tracks8-12-16}. The dependences of average particle oscillation amplitude on the frequency and particle charge are shown in Fig.~\ref{ampl1}. The physical reason of the nonmonotonic decay of some dependences on frequency is possibly related to their vecinity to the resonance frequencies.

\section{Conclusions}\label{concl}
The trap design we have used is currently under test, but preliminary data suggest that it can be used to levitate and detect different aerosol species. We consider using such trap geometry in ultrahigh vacuum conditions. The map of the trap potential coupled with numerical simulations helped us identify extended regions of stable trapping, which extend as far as near the vicinity of the trap electrodes. Such a trap design exhibits an extended region where the trapping field almost vanishes. Microparticle dynamics was investigated by menas of mathematical simulations. Different programming codes are tested, based especially on C++ and Python. A trapped particle microplasma results (very similar to a dusty plasma, which is of great interest for astrophysics), consisting of tens up to thousands of particles. Such a setup is also suited to study and illustrate the appearance of ordered structures and crystal like formations.

As revealed by the experimental data that map the a.c. field within the traps we have investigated coupled with the numerical simulations we have performed, charged particles spend relatively little time in the high RF electric fields area, which results in longer confining times and higher stability. As experimental work  with the trap is currently undergoing, we have used numerical simulations and maps of the trap potential to compare the 16-pole trap with other designs we have tested, such as the 8-electrode and the 12-electrode trap. Regions of stable confinement have been identified, depending on the trap and experimental parameters. These traps are the subject of a different paper that will be published.   

Preliminary results show that thread-like formations of microparticles (dusty plasma) arranged both as planar and volume structures can be observed. These stable structures tend to align with the $z$ component of the radial field. These formations were not disposed on the trap axis and many of them are located rather in the vicinity of the trap electrodes. We are able to report stable confinement for hours and even more. Experimental data shows good agreement with numerical simulation results.

It is our intention to use such traps to confine aerosols and even nanoparticles and we consider they are excellent tools to perform environment studies on presence of polluting agents in the atmosphere and troposphere, coupled with other known techniques such as Lidar. 

\section{Acknowledgements}
\indent
The authors would like to acknowledge support provided by the Ministery of Education, Research and Inovation from Romania (ANCS-National Agency for Scientific Research), contracts PN09.39.03.01 and Ideas Project PN-II-ID-PCE-2011-3-0958, Contract 90/2011. 

Mathematical model and the Brownian dynamics simulations have been carried out in the Joint Institute for High Temperatures from Moscow, Russian Academy of Sciences (RAS), under financial support by the Russian Foundation for Basic Research (grant No. 15-08-02835).

\section{References}
%\printbibliography
 
\bibliographystyle{unsrt}
\bibliography{Multipole}

\begin{thebibliography}{10}

\bibitem{Fortov2007}
Vladimir~E. Fortov, Igor~T. Iakubov, and Alexey~G. Khrapak.
\newblock {\em Physics of Strongly Coupled Plasma}.
\newblock International Series of Monographs on Physics. Clarendon Press,
  Oxford, 2007.

\bibitem{Tsyto2008}
Vadim~N. Tsytovich, Gregory~E. Morfill, Sergey~V. Vladimirov, and Hubertus~M.
  Thomas.
\newblock {\em Elementary Physics of Complex Plasmas}, volume 731 of {\em
  Lecture Notes in Physics}.
\newblock Springer, Berlin Heidelberg, 1st edition, 2008.

\bibitem{Bonitz2014}
Michael Bonitz, Jose Lopez, Kurt Becker, and Hauke Thomsen, editors.
\newblock {\em Complex Plasmas: Scientific challenges and Technological
  Opportunities}, volume~82 of {\em Springer Series on Atomic, Optical and
  Plasma Physics}.
\newblock Springer, Springer International Publishing Switzerland, 1st edition,
  2014.

\bibitem{Campa2014}
A.~Campa, T.~Dauxois, D.~Fanelli, and S.~Ruffo.
\newblock {\em Physics of Long-Range Interacting Systems}.
\newblock Oxford University Press, 2014.

\bibitem{Shukla2002}
Padma~Kant Shukla and A.~A. Mamun.
\newblock {\em Introduction to Dusty Plasma Physics}.
\newblock IOP Publishing Ltd, 2002.

\bibitem{Fortov2010}
Vladimir~E. Fortov and Gregor~E. Morfill, editors.
\newblock {\em Complex and Dusty Plasmas: From Laboratory to Space}.
\newblock CRC Press, 2010.

\bibitem{Fortov2011}
Vladimir~E. Fortov.
\newblock {\em Extreme States of Matter}.
\newblock Springer-Verlag, 2011.

\bibitem{Popel2011}
S.~I. Popel, S.~I. Kopnin, M.~Y. Yu, J.~X. Ma, and Feng Huang.
\newblock The effect of microscopic charged particulates in space weather.
\newblock {\em Journal of Physics D: Applied Physics}, 44:174036, 2011.

\bibitem{Morfill2009}
Gregor~E. Morfill and Alexei~V. Ivlev.
\newblock Complex plasmas: An interdisciplinary research field.
\newblock {\em Rev. Mod. Phys.}, 81:1353--1404, 2009.

\bibitem{Bonitz2010a}
Michael Bonitz, Patrick Ludwig, and Norman Horing, editors.
\newblock {\em Introduction to Complex Plasmas}, volume~59 of {\em Springer
  Series on Atomic, Optical and Plasma Physics}.
\newblock Springer-Verlag, Berlin Heidelberg, 1st edition, 2010.

\bibitem{Thomas1994}
H.~Thomas, G.~E. Morfill, V.~V. Demmel, J.~Goree, B.~Feuerbacher, and
  D.~M{\"o}hlmann.
\newblock Plasma crystal: Coulomb crystallization in a dusty plasma.
\newblock {\em Phys. Rev. Lett.}, 73:652--655, 1994.

\bibitem{Schlipf1996}
Stefan Schlipf.
\newblock From a single ion to a mesoscopic system: Crystalization of ions in
  paul traps.
\newblock In Alain Aspect, W.~Barletta, and R.~Bonifacio, editors, {\em
  Coherent and Collective Interactions of Particles and Radiation Beams},
  volume 131 of {\em Proceedings of the International School of Physics Enrico
  Fermi}, 1996.

\bibitem{Dubin1999}
Daniel H.~E. Dubin and T.~M. O'Neil.
\newblock Trapped nonneutral plasmas, liquids, and crystals (the thermal
  equilibrium states).
\newblock {\em Rev. Mod. Phys.}, 71:87--172, 1999.

\bibitem{Tsyto2007}
V.~N. Tsytovich, G.~E. Morfill, V.~E. Fortov, N.~G. Gusein-Zade, B.~A. Klumov,
  and S.~V. Vladimirov.
\newblock From plasma crystals and helical structures towards inorganic living
  matter.
\newblock {\em New J. Phys.}, 9:263--, 2007.

\bibitem{Vladimirov2005}
S.~V. Vladimirov, K.~Kostrikov, and A.~A. Samarian.
\newblock {\em Physics and Applications of complex Plasmas}.
\newblock Imperial College Press, 2005.

\bibitem{Chen2007}
Goong Chen, David~A. Church, Berthold-Georg Englert, Carsten Henkel, Bernd
  Rohwedder, Marlan~O. Scully, and M.~Suhail Zubairy.
\newblock {\em Quantum Computing Devices: Principles, Design, and Analysis}.
\newblock CRC Press, 2007.

\bibitem{Werth2009}
G{\"u}nther Werth, Viorica~N. Gheorghe, and Fouad~G. Major.
\newblock {\em Charged Particle Traps II: Applications}, volume~54 of {\em
  Springer Series on Atomic, Optical and Plasma Physics}.
\newblock Springer-Verlag, Berlin Heidelberg, 1st edition, 2009.

\bibitem{Paul1990}
Wolfgang Paul.
\newblock Electromagnetic traps for charged and neutral particles.
\newblock {\em Rev. Mod. Phys.}, 62:531--540, 1990.

\bibitem{Major2005}
Fouad~G. Major, Viorica~N. Gheorghe, and G{\"u}nther Werth.
\newblock {\em Charged Particle Traps: Physics and Techniques of Charged
  Particle Field Confinement}, volume~37 of {\em Springer Series on Atomic,
  Optical and Plasma Physics}.
\newblock Springer-Verlag, Berlin Heidelberg, 1st edition, 2005.

\bibitem{Quint2014}
Wolfgang Quint and Manuel Vogel, editors.
\newblock {\em Fundamental Physics in Particle Traps}, volume 256 of {\em
  Springer Tracts in Modern Physics}.
\newblock Springer, Springer-Verlag, Berlin Heidelberg, 2014.

\bibitem{Wuerker1959}
Ralph~F. Wuerker, Haywood Shelton, and Robert~V. Langmuir.
\newblock Electrodynamic containment of charged particles.
\newblock {\em J. Appl. Phys}, 30:342--349, 1959.

\bibitem{Winter1991}
H.~Winter and H.~W. Ortjohann.
\newblock Simple demonstration of storing macroscopic particles in a paul trap.
\newblock {\em J. Appl. Phys}, 59:807--813, 1991.

\bibitem{Ghosh1995}
Pradip~K. Ghosh.
\newblock {\em Ion Traps}.
\newblock Clarendon Press, 1995.

\bibitem{Haroche2013}
Serge Haroche and Jean-Michel Raimond.
\newblock {\em Exploring the Quantum: Atoms, Cavities and Photons}.
\newblock Oxford University Press, 2013.

\bibitem{Davidson2001}
Ronald~C. Davidson.
\newblock {\em Physics of Nonneutral Plasmas}.
\newblock Imperial College Press \& World Scientific, 2001.

\bibitem{Tamashiro1999}
M.N. Tamashiro, Yan Levin, and Marcia~C. Barbosa.
\newblock The one-component plasma: a conceptual approach.
\newblock {\em Phys. A}, 268:24--49, 1999.

\bibitem{Ott2014}
T.~Ott, M.~Bonitz, L.~G. Stanton, and M.~S. Murillo.
\newblock Coupling strength in coulomb and yukawa one-component plasmas.
\newblock {\em Phys. Plasmas}, 21:113704, 2014.

\bibitem{Zagoskin2011}
A.~M. Zagoskin.
\newblock {\em Quantum Engineering: Theory and Design of Quantum Coherent
  Structures}.
\newblock Cambridge Univ. Press, 2011.

\bibitem{Kim2005}
Jungsang Kim, S.~Pau, Z.~Ma, H.~R. McLellan, J.~V. Gates, A.~Kornblit, R.~M.
  Jopson, I.~Kang, M.~Dinu, and Richart~E. Slusher.
\newblock System design for large-scale ion trap quantum information processor.
\newblock {\em Quantum Inf. Comput.}, 5:515--537, 2005.

\bibitem{Leibf2007}
Diedrich Leibfried, David~J. Wineland, R.~Brad Blakestad, John~J. Bollinger,
  Joseph Britton, John Chiaverini, R.~J. Epstein, Wayne~M. Itano, John~D. Jost,
  Emanuel Knill, Christopher Langer, Roee Ozeri, Rainer Reichle, Signe
  Seidelin, N.~Shiga, and J.~H. Wesenberg.
\newblock Towards scaling up trapped ion quantum information processing.
\newblock {\em Hyperfine Interactions}, 174:1--7, 2007.

\bibitem{Riehle2004}
Fritz Riehle.
\newblock {\em Frequency Standards: Basics and Applications}, volume 256.
\newblock Wiley, Wiley-VCH Verlag, Weinheim, 2004.

\bibitem{Major2007}
Fouad~G. Major.
\newblock {\em Quantum Beat: Principles and Applications of Atomic Clocks}.
\newblock Springer, 2007.

\bibitem{March2010}
Raymond~E. March and John F.~J. Todd.
\newblock {\em Practical Aspects of Ion Trap Mass Spectrometry}, volume~5.
\newblock CRC Press, Boca Raton, 1st edition, 2010.

\bibitem{Poli2013}
N.~Poli, C.~W. Oates, P.~Gill, and G.~M. Tino.
\newblock Optical atomic clocks.
\newblock {\em Nuovo Cimento}, 36:555--624, 2013.

\bibitem{Ludlow2015}
Andrew~D. Ludlow, Martin~M. Boyd, Jun Ye, E.~Peik, and P. O. Schmidt.
\newblock Optical atomic clocks.
\newblock {\em Rev. Mod. Phys.}, 87:637--701, 2014.

\bibitem{Trippel2006}
S.~Trippel, J.~Mikosch, R.~Berhane, Rico Otto, Matthias Weidem{\"u}ller, and
  Roland Wester.
\newblock Photodetachment of cold ${OH}^{-}$ in a multipole ion trap.
\newblock {\em Phys. Rev. Lett.}, 97:193003, 2006.

\bibitem{Gerlich2008a}
Dieter Gerlich.
\newblock The study of cold collisions using ion guides and traps.
\newblock In {\em Low Temperatures and Cold Molecules}.

\bibitem{Gerlich2008b}
Dieter Gerlich.
\newblock The production and study of ultra-cold molecular ions.
\newblock In {\em Low Temperatures and Cold Molecules}.

\bibitem{Wester2009}
Roland Wester.
\newblock Radiofrequency multipole traps: tools for spectroscopy and dynamics
  of cold molecular ions.
\newblock {\em J. Phys. B: At. Mol. Opt. Phys.}, 42:154001, 2009.

\bibitem{March2005}
Raymond~E. March and John F.~J. Todd.
\newblock {\em Quadrupole Ion Trap Mass Spectrometry}, volume 165 of {\em
  Chemical Analysis: A Series of Monographs on Analytical Chemistry and its
  Applications}.
\newblock Wiley, Hoboken, New Jersey, 2nd edition, 2005.

\bibitem{Seinfeld2006}
John~H. Seinfeld and Spyros~N. Pandis.
\newblock {\em Atmospheric Chemistry and Physics: From Air Pollution to Climate
  Change}.
\newblock Wiley, 2006.

\bibitem{IPCC}
Ipcc (intergovernmental panel on climate change) wgi fourth assessment report.
\newblock Technical report, 2007.

\bibitem{Kirch2008}
Thomas~W. Kirchstetter, Jeffery Aguiar, Shaheen Tonse, David Fairley, and
  T.~Novakov.
\newblock Black carbon concentrations and diesel vehicle emission factors
  derived from coefficient of haze measurements in california: 1967–2003.
\newblock {\em Atmospheric Environment}, 42:480--491, 2008.

\bibitem{Davidovits2008}
Paul Davidovits.
\newblock The spectroscopy and dynamics of microparticles.
\newblock {\em Faraday Discussions}, 137:425--430, 2008.

\bibitem{Lebedev2012}
Albert~T. Lebedev, editor.
\newblock {\em Comprehensive Environmental Mass Spectrometry}.
\newblock ILM Publications, 2012.

\bibitem{Nash2006}
David~G. Nash, Tomas Baer, and Murray~V. Johnston.
\newblock Aerosol mass spectrometry: An introductory review.
\newblock {\em Int. J. Mass Spectrometry}, 258:2--12, 2006.

\bibitem{Wang2006}
Shenyi Wang, Christopher~A. Zordan, and Murray~V. Johnston.
\newblock Chemical characterization of individual, airborne sub-10 nm particles
  and molecules.
\newblock {\em Anal. Chem.}, 78:1750--1754, 2006.

\bibitem{Pandis1995}
Spyros~N. Pandis, Anthony~S. Wexler, and John~S. Seinfeld.
\newblock Dynamics of tropospheric aerosols.
\newblock {\em J. Phys. Chem.}, 99:9646--9659, 1995.

\bibitem{Kulkarni2011}
Pramod Kulkarni, Paul~A. Baron, and Klaus Willeke, editors.
\newblock {\em Aerosol Measurement: Principles, Techniques and Applications}.
\newblock Wiley, 2011.

\bibitem{IRSST}
Irsst report on health effects of nanoparticles.
\newblock Technical report, Institute de recherche Robert-Sauv`{e} en sante`{e}
  et en securit`{e} de travail, 08 2006.

\bibitem{Seo2003}
Sung~Cheol Seo, Seung~Kyun Hong, and Doo~Wan Boo.
\newblock Single nanoparticle ion trap (snit): A novel tool for studying
  in-situ dynamics of single nanoparticles.
\newblock {\em Bull. Korean Chemical Society}, 24:552--554, 2003.

\bibitem{EPA2012}
EPA.
\newblock Particle pollution and health.
\newblock
  \url{http://www3.epa.gov/airquality/particlepollution/2012/decfshealth.pdf},
  2012.

\bibitem{Gerlich1992}
Dieter Gerlich.
\newblock Inhomogeneous electrical radio-frequency fields: A versatile tool for
  the study of processes with slow ions.
\newblock {\em Adv. Chem. Phys.}, 82:1--176, 1992.

\bibitem{Gerlich2003}
Dieter Gerlich.
\newblock Molecular ions and nanoparticles in rf and ac traps.
\newblock {\em Hyperfine Interactions}, 146.

\bibitem{Otto2009}
R.~Otto, P.~Hlavenka, S.~Trippel, J.~Mikosch, K.~Singer, M.~Weidem{\"u}ller,
  and Roland Wester.
\newblock How can a 22-pole ion trap exhibit ten local minima in the effective
  potential?
\newblock {\em J. Phys. B: At. Mol. Opt. Phys.}, 42:154007, 2009.

\bibitem{Trevi07}
Adam~J. Trevitt, Philip~J. Wearne, and Evan~J. Bieske.
\newblock Calibration of a quadrupole ion trap for particle mass spectrometry.
\newblock {\em Int. J. Mass Spectrometry}, 262:241--246, 2007.

\bibitem{Nie2008}
Zongxiu Nie, Fenping Cui, Minglee Chu, Chung-Hsuan Chen, Huan-Cheng Chang, and
  Yong Cai.
\newblock Calibration of a frequency-scan quadrupole ion trap mass spectrometer
  for microparticle mass analysis.
\newblock {\em Int. J. Mass Spectr.}, 270:8--15, 2008.

\bibitem{Trevi09}
Adam~J. Trevitt, Philip~J. Wearne, and Evan~J. Bieske.
\newblock Coalescence of levitated polystyrene microspheres.
\newblock {\em J. Aerosol Science}, 40:431--438, 2009.

\bibitem{Vogel2013}
A.~L. Vogel, M.~{\"A}ij{\"a}l{\"a}, M.~Br̈{\"u}ggemann, M.~Ehn, H.~Junninen,
  T.~Pet{\"a}j{\"a}, D.~R. Worsnop, M.~Kulmala, J.~Williams, and T.~Hoffmann.
\newblock Online atmospheric pressure chemical ionization ion trap mass
  spectrometry (apci-it-ms$^n$ ) for measuring organic acids in concentrated
  bulk aerosol – a laboratory and field study.
\newblock {\em Atmos. Meas. Tech.}, 4:431--443.

\bibitem{Trevi2006}
Adam~J. Trevitt.
\newblock {\em Ion Trap Studies of Single Microparticles: Optical Resonances
  and Mass Spectroscopy}.
\newblock PhD thesis, School of Chemistry, The University of Melbourne, 2006.

\bibitem{Kurten2007}
Andreas K{\"u}rten, Joachim Curtius, Frank Helleis, Edward~R. Lovejoy, and
  Stephan Borrmann.
\newblock Development and characterization of an ion trap mass spectrometer for
  the on-line chemical analysis of atmospheric aerosol particles.
\newblock {\em Int. J. Mass Spectr.}, 265:30--39, 2007.

\bibitem{Smith2008}
Trevor~A. Smith, Adam~J. Trevitt, Philip~J. Wearne, Evan~J. Bieske, Lachlan~J.
  McKimmie, and Damian~K. Bird.
\newblock Morphology-dependent resonance emission from individual micron sized
  particles.
\newblock In M.~N. Berberan-Santos, editor, {\em Fluorescence of
  Supermolecules, Polymers, and Nanosystems}, volume~4 of {\em Springer Series
  on Fluorescence}, pages 415--429, Berlin Heidelberg, 2008. Springer.

\bibitem{Izmailov1995}
Alexander~F. Izmailov, Stephen Arnold, Stephen Holler, and Allan~S. Myerson.
\newblock Microparticle driven by parametric and random forces: Theory and
  experiment.
\newblock {\em Phys. Rev. E}, 52:1325--1332, 1995.

\bibitem{Vasilyak2013}
Leonid~M. Vasilyak, Vladimir~I. Vladimirov, Lidiya~V. Deputatova, Dmitriy~S.
  Lapitsky, Vladimir~I. Molotkov, Vladimir~Yakovlevich Pecherkin, Vladimir~S.
  Filinov, and Vladimir~E. Fortov.
\newblock Coulomb stable structures of charged dust particles in a dynamical
  trap at atmospheric pressure in air.
\newblock {\em New Journal of Physics}, 15:043047, 2013.

\bibitem{Lapitsky2015}
D.~S. Lapitsky.
\newblock Effective forces and pseudopotential wells and barriers in the linear
  paul trap.
\newblock {\em J. Phys: Conference Series}, 653:012130, 2015.

\bibitem{Gheorghe98}
Viorica~N. Gheorghe, Liviu Giurgiu, Ovidiu Stoican, Drago{\c s} Cacicovschi,
  Radu Molnar, and Bogdan Mihalcea.
\newblock Ordered structures in a variable length {AC} trap.
\newblock {\em Acta Physica Polonica A}, 93:625--629, 1998.

\bibitem{Stoican2008}
Ovidiu~S. Stoican, Lauren{\c t}iu~C. Dinc{\u a}, Gina Vi{\c s}an, and {\c
  S}tefan R{\u a}dan.
\newblock Acoustic detection of the parametrical resonance effect for a
  one-component microplasma consisting of the charged microparticles stored in
  the electrodynamic traps.
\newblock {\em J. Opt. Adv. Mat.}, 10:1988--2000, 2008.

\bibitem{Mihalcea2008}
Bogdan~M. Mihalcea, Gina T.~Vi\c san, Liviu~C. Giurgiu, and {\c S}tefan R{\u
  a}dan.
\newblock Optimization of ion trap geometries and of the signal to noise ratio
  for high resolution spectroscopy.
\newblock {\em J. Opt. Adv. Mat.}, 10:1994--1998, 2008.

\bibitem{Visan2013}
Gina~Vi\c san and Ovidiu Stoican.
\newblock An experimental setup for the study of the particles stored in an
  electrodynamic linear trap.
\newblock {\em Rom. J. Phys.}, 58:171--180, 2013.

\bibitem{Lapitskiy2015}
D.~S. Lapitskiy, V.~S. Filinov, L.~V. Deputatova, L.~M. Vasilyak, V.~I.
  Vladimirov, and V.~Ya. Pecherkin.
\newblock Capture and retention of charged dust particles in electrodynamic
  traps.
\newblock {\em High Temperature}, 53:1--8, 2015.

\bibitem{Guan2011}
Weihua Guan, Sony Joseph, Jae~Hyun Park, Predrag~S. Krsti\'c, and Mark~A. Reed.
\newblock Paul trapping of charged particles in aqueous solution.
\newblock {\em Proc. Natl. Acad. Sci.-PNAS}, 108:9326--9330, 2011.

\bibitem{We2}
V.~S. Filinov, D.~S. Lapitsky, L.~V. Deputatova, L.~M. Vasilyak, V.~I.
  Vladimirov, and O.~A. Sinkevich.
\newblock {\em Contrib. Plasma Phys.}, 52:66--69, 2012.

\bibitem{Skeel2002}
R.~D. Skeel and J.~A. Izaguirre.
\newblock {\em Molecular Physics}, 100:3885--3891, 2002.

\end{thebibliography}

\end{document}